\documentclass[aps,prl,twocolumn,amsmath,amssymb,floatfix,nofootinbib]{revtex4-1}

\newcommand{\nhat}{\hat{ \mathbf{n}}}

\newcommand{\lvec}{\mathbf{l}}

\usepackage{graphicx}
\usepackage{dcolumn}
\usepackage{bm}
\usepackage[usenames, dvipsnames]{color}

\graphicspath{ {figures/} }

\newcommand{\zsource}{z_\mathrm{s}}
\newcommand{\beelm}{B_{\ell m}}
\newcommand{\omegalm}{\omega_{\ell m}}

\newcommand{\be}{\begin{eqnarray}}

\newcommand{\ee}{\end{eqnarray}}

\begin{document}


\title{Establishing the origin of CMB $B$-mode polarization}


\author{Connor Sheere$^{1,2}$}
\author{Alexander van Engelen$^1$}
\author{P. Daniel Meerburg$^1$}
\author{Joel Meyers$^1$}
\affiliation{$^1$CITA, University of Toronto, 60 St. George Street, Toronto, Canada}
\affiliation{$^2$Department of Physics, McGill University, 3600 rue University, Montr\'eal, Canada}


\date{\today}

\begin{abstract}
Primordial gravitational waves leave a characteristic imprint on the cosmic microwave background (CMB) in the form of $B$-mode polarization.  Photons  are also deflected by large scale gravitational waves which intervene between the source screen and our telescopes, resulting in curl-type gravitational lensing.  Gravitational waves present at the epoch of reionization contribute to both effects, thereby leading to a non-vanishing cross-correlation between $B$-mode polarization and curl lensing of the CMB.  Observing such a cross correlation would be very strong evidence that an observation of $B$-mode polarization was due to the presence of large scale gravitational waves, as opposed to astrophysical foregrounds or experimental systematic effects. We study the cross-correlation across a wide range of source redshifts and show that a post-SKA experiment aimed to map out the 21-cm sky between $15 \leq z \leq 30$ could rule out non-zero cross-correlation at high significance  for $r \geq 0.01$. 
\end{abstract}


\maketitle

\section{Introduction}
Primordial gravitational waves are a key observational target for ongoing and upcoming cosmological surveys. Since direct detection of primordial gravitational waves is likely out of reach for the foreseeable future, we must rely on indirect methods. The most promising probe is in the polarization of the cosmic microwave background (CMB), which is generated by Thomson scattering on free electrons in the presence of a local temperature quadrupole.

On large angular scales, $B$ modes in the polarization field are generated only from scattering on quadrupoles that are sourced by gravitational waves  \cite{Zaldarriaga:1996xe,1997PhRvL..78.2058K,Kamionkowski:2015yta}. 

Currently, our best constraints on this observable 
come from a joint analysis of the BICEP/Keck and \textit{Planck} data \cite{2016PhRvL.116c1302B}, with an upper limit on the ratio of power in gravitational waves to density fluctuations  of  $r<0.07$.  Future CMB surveys are expected to constrain gravitational waves at the level of $\sigma_r \simeq 10^{-3}$ \cite{LiteBird,CLASSexp,COREplus,CMBS4ScienceBook}. If $B$ modes are detected in the CMB on large angular scales, it will be important to confirm that they were sourced by primordial gravitational waves \cite{hirata2012}.

Another probe of primordial gravitational waves is the curl mode of weak gravitational lensing on large scales.  Gravitational waves deflect photons in patterns which contain divergence-free components.   
This leads to a curl part in the distribution of shapes of observed sources, such as galaxy ellipticities or hot and cold spots of the CMB \footnote{In the weak lensing literature this is often denoted the $B$ mode of lensing, but to avoid confusion with $B$ modes in CMB polarization we will follow \cite{CurlyCooray,Namikawa:2014lla} and call these $\omega$ modes.}.  This signal is routinely estimated as a check for systematic contamination in estimates of weak lensing by density fluctuations, which are expected to produce only curl-free deflection on large scales.  However, it has also been considered as a probe of primordial gravitational waves using a variety of lensed sources, such as optical galaxies \cite{Stebbins:1996wx, Dodelson:2003bv, 2010PhRvD..82b3522D}, the CMB \cite{Li:2006si,Namikawa:2014lla}, and intensity mapping surveys such as 21-cm observations \cite{PhysRevLett.108.211301}.  The latter probe, though futuristic, is particularly promising, with the potential to constrain the tensor-to-scalar ratio down to $\sigma_r \sim 10^{-9}$.

\begin{figure*}[t]
\begin{center}
\includegraphics[width = 2\columnwidth]{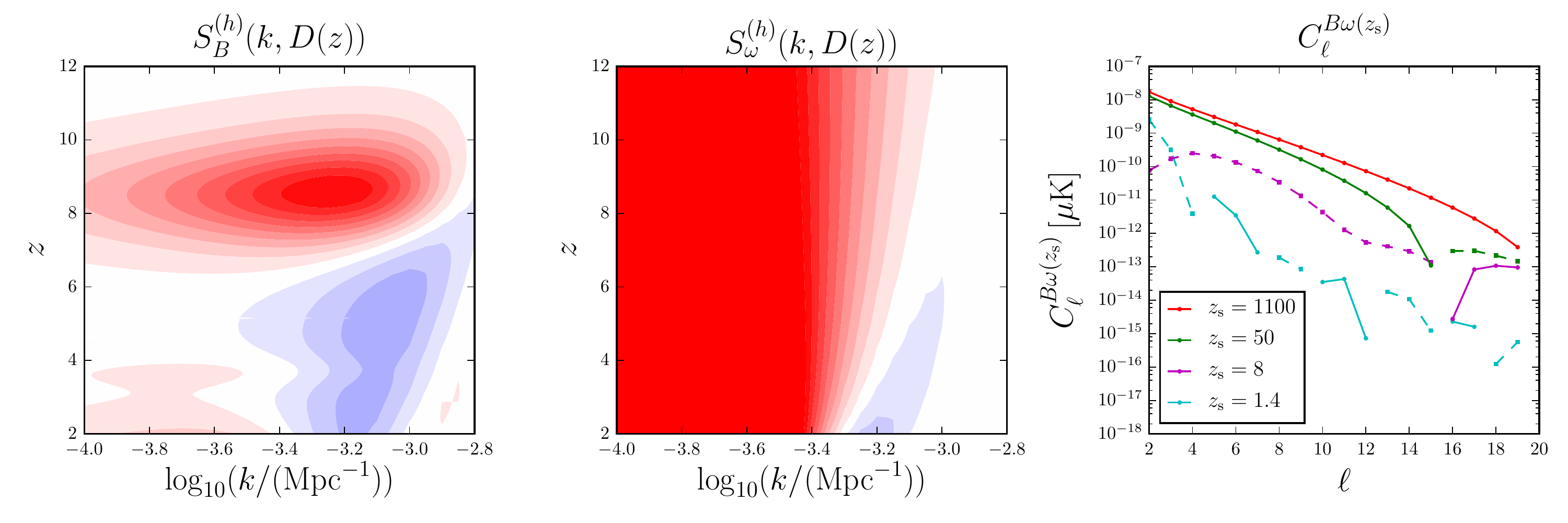}
\caption{\textit{Left:} Polarization source function for gravitational waves yielding $B$ modes from reionization. \textit{Middle:} Source function for gravitational waves yielding curl lensing $\omega$ modes; this is part of the integrand of Eq.~(\ref{eq:curllensingtransfer}) which is performed up to the redshift of the lensing sources.  This panel does not include the metric shear term (Eq.~\ref{eq:metricshear}). \textit{Right:}  Cross power spectra $C_l^{B\omega(\zsource)}$ resulting from the nonzero overlap between these source functions for various source redshifts $\zsource$ (including Eq.~(\ref{eq:metricshear})).  Dashed lines indicate negative values.}
\label{fig:alphazsource}
\end{center}
\end{figure*}

There is a connection between these two observables.  As described by \cite{2010PhRvD..82b3522D}, CMB $B$ modes from reionization, which dominate the $B$-mode signal at multipoles $l < 20$,
are sourced from gravitational waves that yield a temperature quadrupole at the scattering redshift at $z \lesssim 10$, and these same gravitational waves will yield curl lensing deflection $\omega$ for background sources.  Since  $B$ and $\omega$ are both pseudo-scalar fields, the cross-correlation will not vanish due to parity.
As discussed in Refs.~\cite{2010PhRvD..82b3522D, 2014PhRvD..90d3527C}, if detectable  this cross-correlation could be used to confirm or rule out the primordial nature of any claimed detection of gravitational waves from large-scale $B$ modes alone.

Thus far, this correlation has been studied in the context of curl lensing of optical sources at $\zsource \sim 1$ to $2$ \cite{2010PhRvD..82b3522D, 2014PhRvD..90d3527C}.  However, since the  $B$ modes from reionization are generated after $z \sim 10$ we expect that curl lensing reconstructed from higher-redshift screens might yield a larger correlation.  

In this short paper we consider this cross-correlation for high-redshift lensing sources.  
We focus on surveys in a number of redshift ranges at which  nearly full-sky surveys are either in progress, being planned, or being considered.  We consider the CMB at $\zsource \sim 1100$, as well as 21-cm intensity mapping surveys in three redshift regimes: post-reionization around $\zsource \sim 2 $, reionization around $\zsource \sim 9$, and dark ages around $\zsource \sim 50$.

After reviewing the physics of the $B$-mode polarization and $\omega$-mode lensing sourced by gravitational waves, we compute the expected cross power spectra.  We then compute the sensitivity to this cross-correlation for a number of surveys,  focusing on whether they would be useful as a check of the primordial nature of any $B$-mode detection from the CMB alone.

\section{CMB polarization \label{sec:CMBpol}}
Primordial gravitational waves produce CMB fluctuations in temperature and polarization.  Because the temperature and and the $E$-mode polarization suffer from large cosmic variance induced by scalars, the $B$ modes are the cleanest channel in which to search for primordial gravitational waves.
We will start by assuming an isotropic stochastic background of primordial gravitational waves 
\be
\left\langle h^{\lambda}(\mathbf{k}_1) h^{\lambda' }(\mathbf{k}_2) \right\rangle  = \frac{(2\pi)^3}{2} \delta(\mathbf{k}_1 + \mathbf{k}_2) P_h(k)\delta^{\lambda \lambda'}
\ee	
described by a power spectrum
\begin{equation}
P_h(k) = rA_Sk^{-3}\left(\dfrac{k}{k_0}\right)^{n_T}.
\end{equation}
Here, $A_S \sim 10^{-10}$ is the amplitude of primordial scalar fluctuations, $r$ is the tensor-to-scalar ratio, and we will take $n_T\simeq 0$ to study a nearly scale-invariant spectrum. 

Primordial $B$-mode polarization in the CMB can be described in terms of spherical harmonic coefficients, which are directly proportional to the primordial tensor modes 
\be
B_{\ell m} = 4\pi (-i)^{\ell} \int \frac{d^3k}{(2\pi)^3}\sum_{ \pm} {}_{\pm 2}Y^*_{\ell m}(\mathbf{\hat k}) h^\pm(\mathbf{k}) T_{l}^{B}(k).
\ee
Here the transfer functions $T_l^{X}(k)$ are computed by integrating the source functions obtained from a Boltzmann integrator in linear cosmological perturbation theory over conformal time $\eta$ (using the normalization convention of Ref.~\cite{CLASSpol}):
\begin{equation}
\label{eq:poltransfer}
T_{l}^{B}(k) = \sqrt{\dfrac{\pi^2}{8}}\int_{0}^{\eta_0} d\eta S_P^{(h)}(k,\eta) \hat{\mathcal{B}}(k \eta)\dfrac{j_l(k\eta)}{(k\eta)^2}.
\end{equation}
Here,  $\eta_0$ is the conformal time today, $\hat {\mathcal{B}}(x) = 8x + 2x^2 \partial_x^2$ and the tensor ($h$) polarization source function is given by $S_P^{(h)}(k, \eta) = -g(\eta) \Psi(k,\eta)$ \cite{Zaldarriaga:1996xe} with $g(\eta)$ the visibility function for Thomson scattering and $\Psi$ the Newtonian gravitational potential. Integration by parts and a change of variables to $D = \eta_0-\eta$ yields
\be
T_{l}^{B}(k) = \sqrt{\dfrac{\pi^2}{8}}\int_{0}^{D_*} dD S_B^{(h)}(k,D) j_{\ell}(k D)
\ee
with $S_B^{(h)}(k,\eta)$ the $B$-mode source function, given by
\be
S_B^{(h)}(k,\eta) =  -g(\eta)\left( \frac{\Psi(k,\eta)}{k\eta}+\frac{2 \dot{\Psi}(k,\eta)}{k} \right) - 2\dot{g}(\eta) \frac{\Psi(k,\eta)}{k}.\nonumber
\ee
 
\begin{figure}[t]
\begin{center}
\includegraphics[width=\columnwidth]{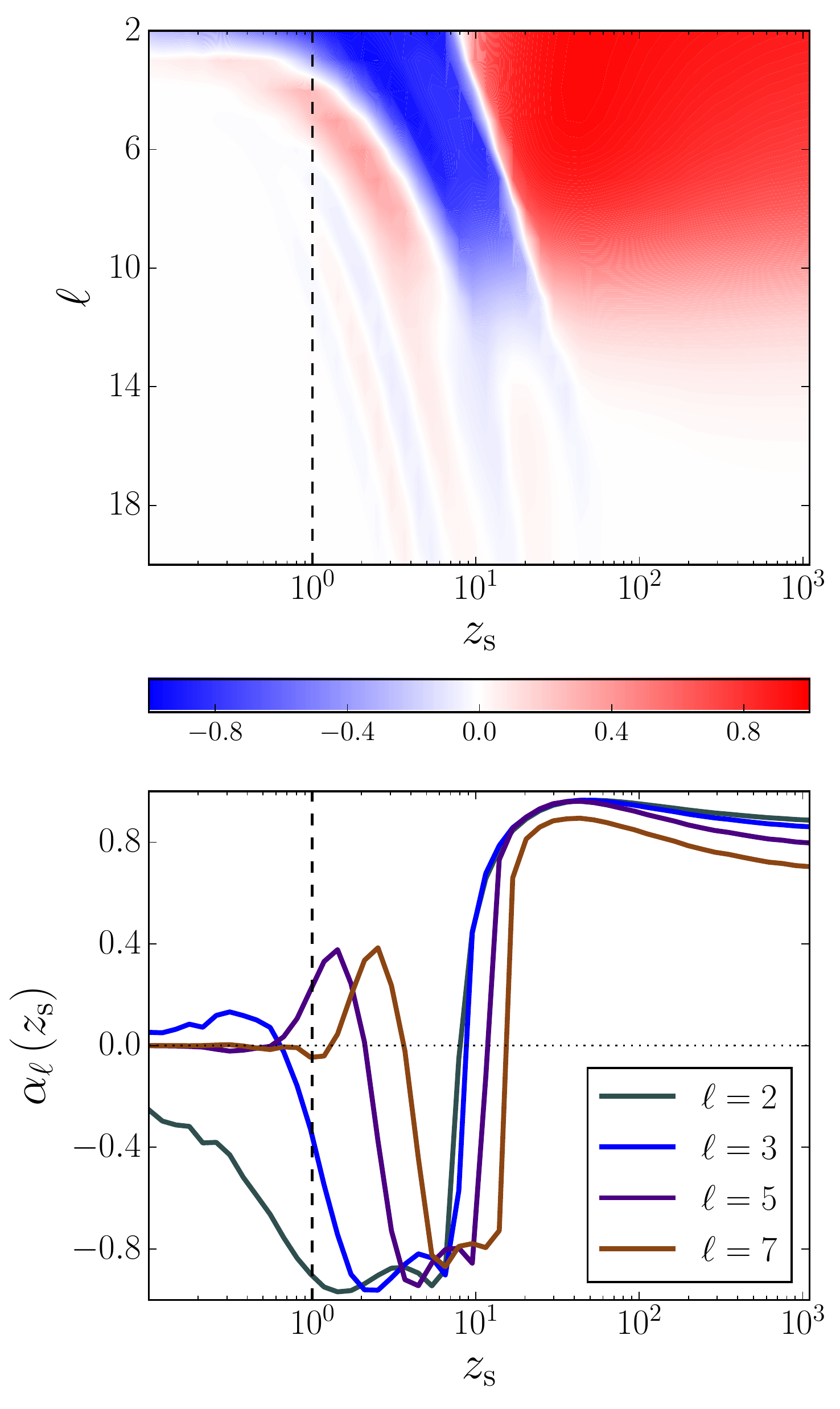}
\caption{\textit{Top:} The cross-correlation coefficient $\alpha_\ell(\zsource)$ (Eq.~(\ref{eq:alphadefinition})) for $\ell \leq 20$ and $0\leq \zsource \leq 1100$. The cross-correlation coefficient is larger at $\zsource \gtrsim 3$ and drops quickly for $\ell \gtrsim 10$. \textit{Bottom:} Cross-correlation coefficient $\alpha_\ell(\zsource)$ as a function of $\zsource$ for several values of $\ell$.  Previous work has focused on sources around $\zsource \sim 1$ (dashed line), but for $\ell > 3$ the correlation is larger for higher-redshift sources, such as those from the high redshift 21-cm and the CMB.}
\label{fig:alpha}
\end{center}
\end{figure}

\section{Curl Lensing}\label{sec:lensing}
The deflection due to gravitational lensing can be decomposed into a curl-free part $\phi(\nhat)$ and a divergence-free part $\Omega(\nhat)$.  Images $\tilde X(\nhat)$ such as the  CMB temperature, the CMB polarization Stokes parameters $ Q(\nhat) \pm i  U(\nhat)$, or  the brightness fluctuations in an intensity mapping survey are remapped according to
\be
X(\nhat) = \tilde X(\nhat + \nabla \phi(\nhat) + \nabla \times \Omega(\nhat)),
\ee
where $(\nabla \times \Omega(\nhat))_i = \epsilon_{ij} \partial_j \Omega$ with $\epsilon_{ij}$ the antisymmetric tensor.
Although curl lensing modes $\Omega$ can be sourced on small scales by lens-lens coupling \cite{Cooray:2002mj, Pratten:2016dsm, Hirata:2003ka},  on large scales, non-zero $\omega$ is expected to be a signature of gravitational waves \cite{Li:2006si}. 
Defining $\omega(\nhat) = -{1 \over 2} \nabla^2 \Omega(\nhat)$,  the curl lensing field $\omega$  can be expanded in spin-weighted spherical harmonics as 
\be
\omega_{\ell m} = 4\pi (-i)^{\ell} \int \frac{d^3k}{(2\pi)^3}\sum_{ \pm} {}_{\pm 2}Y^*_{\ell m}(\mathbf{\hat{k}}) h^\pm(\mathbf{k})T_{\ell}^{\omega(\zsource)}(k) \, .\nonumber \\
\ee
The lensing transfer functions for sources at a given redshift $\zsource$ are (e.g. \cite{2010PhRvD..82b3522D})
 \begin{equation}
T_{\ell}^{\omega(\zsource)}(k) = \sqrt{\dfrac{(\ell+2)!}{(\ell-2)!}}\int_{0}^{D(\zsource)} dD S^{(h)}_{\omega}(k,D) j_{\ell}(kD)
\label{eq:curllensingtransfer}\, ,
\end{equation}
where $S^{(h)}_{\omega}(k,D)	 = T^{(h)}(k, D)/(kD^2)$ and $T^{(h)}(k, D) = 3j_{1}(k(\eta_0-D))/(k(\eta_0-D))$ is an approximate solution  to the wave equation describing the evolution of gravitational waves after re-entry into the horizon. 

There is a correction to the tranfer function above caused by the shearing of the coordinates with respect to physical space. Assuming physical isotropy, this adds a ``metric shear''  \cite{Dodelson:2003bv,Schmidt:2012nw} 
\be
\Delta T_{\ell}^{\omega(\zsource)}(k) &=&  \frac{1}{(\ell+2)(\ell-1)} \sqrt{\dfrac{(\ell+2)!}{(\ell-2)!}}\nonumber \\ && \times\left[\frac{\ell-1}{kD(\zsource)}j_{\ell}(kD(\zsource))-j_{\ell-1}(kD(\zsource))\right]\nonumber \\
&& \times T^{(h)}(k, D(\zsource)) \, .
\label{eq:metricshear}
\ee
We include this term in the analysis that follows. 

\section{Cross-correlation}\label{sec:crosscorr}
The auto- and cross-spectra are computed by correlating the spherical harmonic coefficients, i.e. $\langle X_{\ell m} Y_{\ell' m'}^*\rangle = \delta_{mm'}\delta_{\ell\ell'}C_\ell^{XY}$, where 
\begin{equation}
C_{\ell}^{XY} = \dfrac{2}{\pi}\int{dk k^2 P_h(k)T_{\ell}^{X}(k)T_{\ell}^{Y*}(k)} \, ,
\end{equation} 
with $\{X,Y\} \in\{B,\omega\}$ and the transfer functions are given in the previous sections.  We extract the polarization source functions from the CLASS code \cite{CLASS,CLASSpol}.  The contributions from reionization were isolated by restricting the integral over conformal time $\eta$.  We use parameters consistent with results from the {\it Planck} survey \cite{Ade:2015xua}. 

In the right panel of Fig.~\ref{fig:alphazsource} we show the cross power spectrum $C_l^{B\omega(\zsource)}$ for a set of source redshifts $\zsource$.  This is the predicted signal in linear cosmological perturbation theory in  a universe with gravitational waves. For the CMB case at $\zsource = 1100$, this is the gravitational wave analog of the reionization-generated cross power spectrum $C_{\ell}^{E\phi(\zsource)}$ between polarization $E$ modes and CMB lensing potential $\phi$ modes sourced by density fluctuations computed in Ref.~\cite{Lewis:2011fk}.

In Fig.~\ref{fig:alpha} we show the cross-correlation coefficient, defined as
\begin{equation}
\alpha_{\ell}(\zsource) = \dfrac{C_{\ell}^{B\omega(\zsource)}}{\sqrt{C_{\ell}^{BB}C_{\ell}^{\omega\omega(\zsource)}}} \, .
\label{eq:alphadefinition}
\end{equation}
These two fields are strongly correlated for multipoles $\ell<  20$, particularly for high-redshift sources.

\begin{figure}[t!]
\begin{center}
\includegraphics[width=\columnwidth]{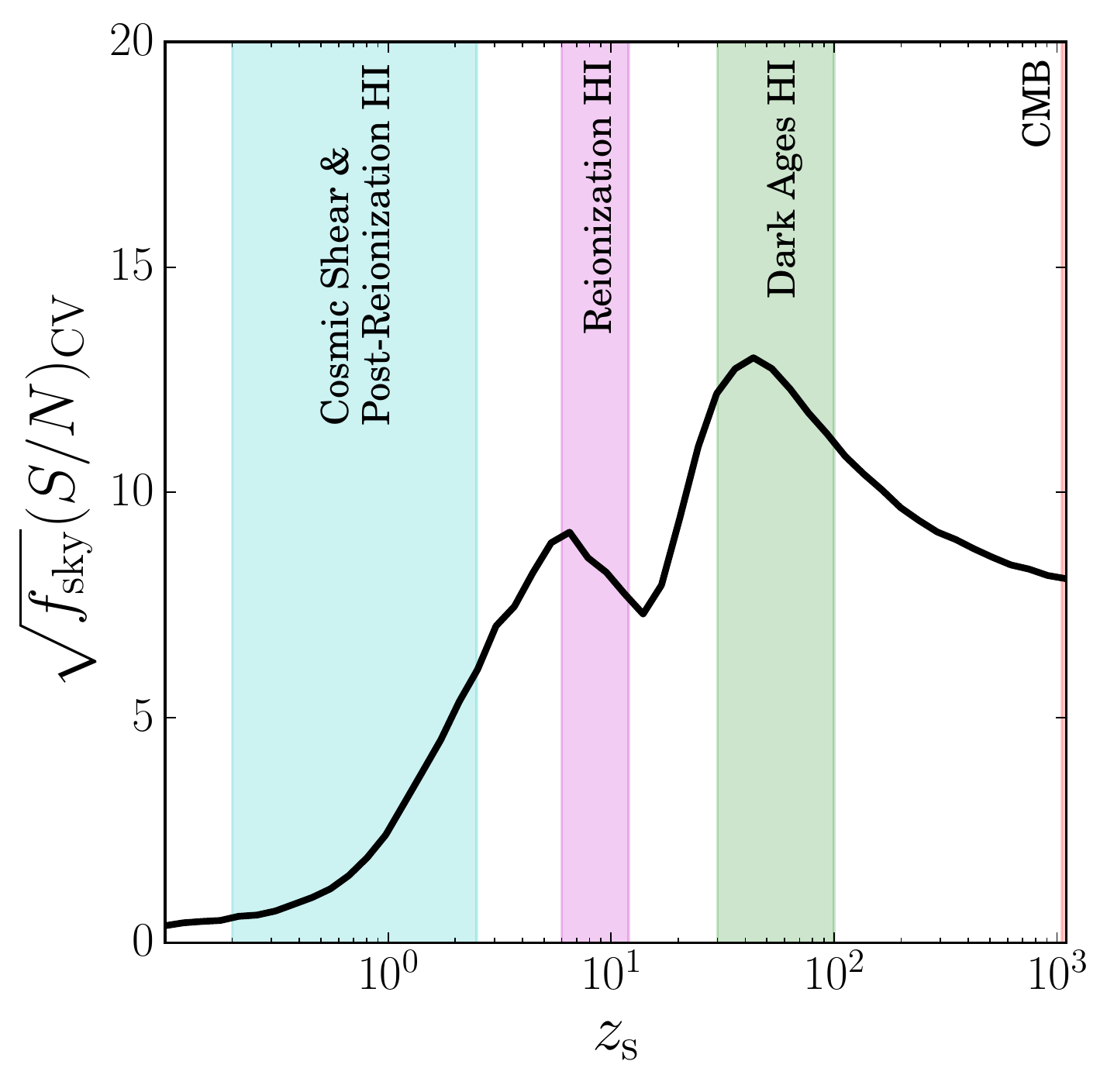}
\caption{Detection significance for correlation in the case of cosmic variance limited observations of both $B$ and $\omega(\zsource)$ as a function of $\zsource$. Colored bands represent eras probed by different cosmological surveys. The plot suggests that the cross-correlation is best observed at higher redshifts, with the maximum signal situated around $\zsource \sim 50$. This curve is independent of the value of the tensor-to-scalar ratio, $r$.  In principle, many redshift bins could be combined to significantly boost the total detectability.}
\label{fig:SN_CV}
\end{center}
\end{figure}

\section{Detectability}\label{sec:SN}

We would like to estimate the cross-correlation between a (perhaps noisy) map of CMB $B$ modes and a reconstructed map of $\omega$ modes at some source redshift $\zsource$.  Assuming full-sky data, for a given value of $\ell$ we could use the $2l+1$ pairs of observed spherical harmonic moments ($\beelm$, $\omegalm$) to form the  cross-correlation statistic\footnote{For small values of $2\ell+1$, this estimate of the observed correlation coefficient is biased, and should be corrected by using $\hat\rho' = \hat\rho\left(1+\frac{(1-\hat\rho^2)}{4\ell+2}\right)$ instead~\cite{Fisher:1915ab}.  In addition, for nonzero experimental noise the estimate $\hat \rho_\ell$ is not an unbiased estimator for the predicted correlation $\alpha_\ell$ (Eq.~(\ref{eq:alphadefinition})), with $|\langle\hat\rho_\ell\rangle| < |\alpha_\ell|$.  However, it can be used to rule out the case of no correlation.}
\be
\hat{\rho}_\ell = \frac{\sum_m \beelm \omegalm^* }{ \sqrt{\sum_m |\beelm|^2 \sum_{m^\prime} |\omega_{\ell m^\prime}|^2}} \, .
\label{eq:rhohat}
\ee

The  estimator $\hat{\rho}_\ell$ can have a highly non-Gaussian distribution.  For instance, in the limit where the experimental and reconstruction noises are very small
and where the cross-correlation  $\hat{\rho}_\ell$ becomes very close to 1, the values of $\beelm$ and $\omegalm$ would become exactly equal (up to an overall multiplicative constant).  The cross correlation would then be detectable at extremely high significance, even though the number of independent  modes for a given $\ell$, namely $2\ell + 1$, might not be large.   In other words,  this cross-correlation could in principle be detectable without being limited by cosmic variance.

To render the cross-correlation estimate more Gaussian, one can use the Fisher transformation \cite{Fisher:1915ab,Fisher:1921bc}, given by
\be
\hat y_\ell = {1 \over 2}\ln \left({1 + \hat \rho_\ell \over 1 - \hat \rho_\ell}\right) = \operatorname{atanh}(\hat\rho_\ell) \, .
\ee
For $2l+1$ independent samples drawn from a bivariate Gaussian distribution in  $(\beelm, \omegalm)$, this transformed estimator has a  nearly Gaussian distribution, with mean 
\be
\langle \hat{y}_\ell \rangle = 
        \operatorname{atanh}\left(\frac{C_\ell^{B \omega(\zsource)} }
          { \sqrt{\left(C_\ell^{BB} + N_\ell^{BB}\right)\left(C_\ell^{\omega\omega(\zsource)} + N_\ell^{\omega\omega(\zsource)}\right)}} \right)\nonumber \\ \, .          
\label{eq:rhohatmean}
\ee
and variance $\sigma_{\hat y_\ell}^2 = 1/(2\ell - 2)$.\footnote{We note that although cosmological $B$ and $\omega$ modes are expected to be Gaussian, this will not formally be the case for the noise in the fields, especially for the $\omega$ field which is generally obtained using quadratic estimators.  We also neglect non-Gaussianity in the 21-cm field itself \cite{Lu:2007pk}.}
Here, $C_\ell^{BB}$ and $C_\ell^{\omega \omega(\zsource)}$ are the cosmological contributions to the variances in the $\beelm$ and $\omegalm$ while the $N_\ell^{BB}$ and $N_\ell^{\omega \omega(\zsource)}$ are the contributions from instrumental and reconstruction noise. 

We take $N_\ell^{BB}$ to be a constant on the scales of interest, parameterized by a pixel noise level $N_\ell^{BB} = 2(\Delta_T)^2$. For   $N_\ell^{\omega \omega(\zsource)}$
we work in the flat-sky approximation\footnote{This is a reasonable approximation because the curl lensing noise 
is nearly constant at large angular scales~\cite{Hu:2001kj,CurlyCooray}. } and use the quadratic estimator of Ref.~\cite{Hu:2001kj,CurlyCooray} for the CMB, generalized to an intensity mapping survey with multiple redshift screens by Ref.~\cite{2006ApJ...653..922Z}:
\be
\left(N_{\ell = |\lvec|}^{\omega \omega(\zsource)}\right)^{-1} &=& \sum_{j = j_{\rm min}}^{j_{\rm max}} \frac{4}{|\lvec|^4} \int \frac{d^2 \lvec^\prime}{(2 \pi)^2} f_\beta^j(\lvec^\prime, \lvec - \lvec^\prime) F_\beta^j(\lvec^\prime, \lvec - \lvec^\prime), \nonumber \\
\label{eq:lensingnoise}
\ee
where $j$ indexes radial modes.
For the CMB, which has only the $j = 0$ term, we use the $\beta = EB$ quadratic estimator, with $f_\beta^j(\lvec_1, \lvec_2) = ((\lvec_1 + \lvec_2) \times \lvec_1) C_{\lvec_1}^{EE} \sin 2\varphi_{\lvec_1,\lvec_2}$ and $F_\beta^j(\lvec_1, \lvec_2) = f_\beta^j(\lvec_1, \lvec_2) / C_{|\lvec_1|}^{EE,\mathrm{tot}}C_{|\lvec_2|}^{BB,\mathrm{tot}}$, where the $C_{|\lvec|}^{XX,\mathrm{tot}}$ denote the total power spectra including noise.  For intensity mapping surveys, with $\beta = II$, we take $f_\beta^j(\lvec_1, \lvec_2) = (\lvec_1 + \lvec_2) \times (\lvec_1 C_{\lvec_1}^{II} + \lvec_2 C_{\lvec_2}^{II})$ and $F_\beta^j(\lvec_1, \lvec_2) = f_\beta^j(\lvec_1, \lvec_2) / 2 C_{|\lvec_1|}^{II,\mathrm{tot}}C_{|\lvec_2|}^{II,\mathrm{tot}}$.

\begin{figure}[t!]
\begin{center}
\includegraphics[width=\columnwidth]{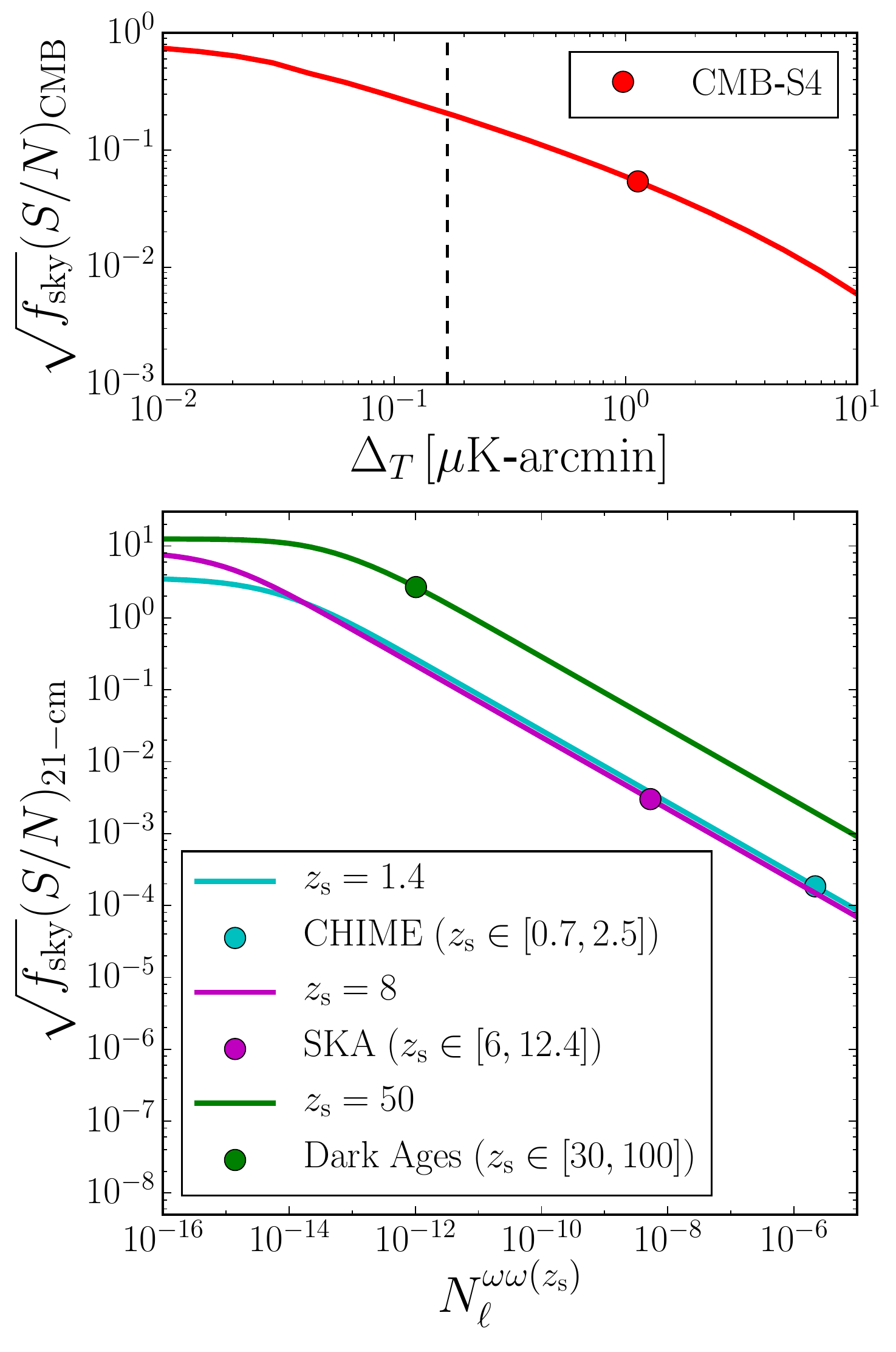}
\caption{\textit{Top:} Significance to reject the case of no correlation using only the CMB, as a function of the noise $\Delta_T$, assuming $r = 0.01$. \ The dashed curve indicates the noise level at which curl and convergence modes might  not be distinguishable~\cite{Hirata:2003ka}. \textit{Bottom:} Significance for 21-cm surveys as a function of the curl-lensing reconstruction noise $N_{\ell}^{\omega \omega(\zsource)}$, assuming $r = 0.01$ and CMB noise $\Delta_T = 0.25$ $\mu$K-arcmin for several ranges of source redshifts. 
Using multiple redshift screens can serve to reduce the noise $N_{\ell}^{\omega \omega(\zsource)}$; for the three forecasted points shown, we sum over the given redshift range using the experimental configurations described in the text.}
\label{fig:SN_CMB_21}
\end{center}
\end{figure}

One can then form the chi-square statistic with respect to the expectation for no correlation, using $\chi^2  = \sum_\ell(\hat y_\ell / \sigma_{\hat y_\ell})^2$.  For forecasting results for future data, we average over datasets, yielding
\be
\langle \chi^2\rangle = \sum_\ell {(2\ell - 2)}  \langle \hat y_\ell \rangle^2\, ,
\ee
with $\langle \hat y_\ell\rangle$ given by Eq.~(\ref{eq:rhohatmean}). We compute  $\langle \chi^2\rangle$ at a variety of source redshifts and for a range of noise levels $N_\ell^{BB}$ and $N_\ell^{\omega\omega(\zsource)}$.  We report the significance with which we can reject the null hypothesis of no cross-correlation $\sqrt{\chi^2}$, which we loosely refer to as $(S/N)$ in Figs.~\ref{fig:SN_CV} and \ref{fig:SN_CMB_21}. \vspace{0.7em}

\noindent

\noindent {\bf CMB ($\zsource\sim 1100$):}
For curl modes reconstructed using the CMB, we use the EB estimator, including iterative delensing of the scalar-induced $B$ modes \cite{Smith:2010gu}.  Although yielding a lower noise level, using polarization rather than temperature for reconstruction actually reduces the overall $\omega$ signal by a factor of approximately 2  \cite{Dai:2013nda}, which we have accounted for.
In the top panel of Fig.~\ref{fig:SN_CMB_21}, we show the significance with which the case of no correlation can be rejected, as a function of the survey noise in $\mu$K-arcmin.  We have assumed that the  $\omega$  map and the $B$ map  are obtained from surveys  with the same noise level; although the former can be obtained with  a ground-based survey such as CMB-S4 \cite{CMBS4ScienceBook}, the latter will require  a nearly full-sky  experiment to measure large scales (e.g. LiteBIRD \cite{LiteBird} or  CORE \cite{COREplus} from space; CLASS \cite{CLASSexp} from the ground).    
For very low noise levels of $\Delta_T \sim 0.17$ $\mu$K-arcmin, it might become difficult to distinguish curl modes from divergence modes  \cite{Hirata:2003ka}.  Even for noise levels    $\Delta_T \sim 10^{-2}$ $\mu$K-arcmin, the cross-correlation is not detectable for $r = 0.01$. 
\vspace{0.7em}

For our 21-cm forecasts, we assume that the CMB is mapped with $\Delta_T = 0.25$ $\mu$K-arcmin and show results in Fig.~\ref{fig:SN_CMB_21} as a function of the curl lensing noise $N_\ell^{\omega\omega(\zsource)}$ at various redshifts.  The lensing reconstruction noise is in general a complicated function of antenna noise, configuration, baseline, frequency resolution, and bandwidth, but we simply assume a constant reconstruction noise at large scales in order to determine what is required to achieve a null rejection.  We also show forecasts for a few specific experimental configurations described below. \vspace{0.7em}

\noindent
{\bf Dark Ages ($30 \leq \zsource \leq 100$):} 
The 21-cm signal from the dark ages can be computed analytically \cite{Loeb_2004,Bharadwaj_2004} and we will consider experimental limitations as in Ref.~\cite{Chen:2016zuu}. We assume that foreground cleaning will remove large-scale modes along the line of sight, 
which we take into account by setting $j_{\rm min} \geq 3$ as in \cite{2006ApJ...653..922Z,Pourtsidou21032014}. We assume a 3 MHz bandwidth and a frequency resolution of $0.01$ MHz. Note that in practice the frequency resolution can easily be improved, however lensing reconstruction noise is typically most limited by the experimental baseline.  We estimate the minimal baseline required to  obtain a 3 sigma rejection of the null hypothesis given a tensor-to-scalar ratio $r = 0.01$ and find that a baseline of $\sim 100$~km would suffice. Note that all we need is the lensing modes for $2 \leq \ell \leq 20$. 
Our forecast is more realistic than Ref.~\cite{PhysRevLett.108.211301}, where the experiment under consideration was cosmic variance limited to $\ell = 10^6$, which at $z = 100$ translates to a baseline of $\sim 5000$~km.  
The points in the bottom panel of Fig.~\ref{fig:SN_CMB_21} are obtained  by combining all  independent redshifts to estimate the total noise reconstruction and assuming an average signal within our window.   \vspace{0.7em}

\noindent
{\bf Reionization ($6 \leq \zsource \leq 12$):} For a survey probing the reionization era \cite{peneorlensing}, we consider a low-SKA type experiment to reconstruct the lensing noise through Eq.~\eqref{eq:lensingnoise}. Basic experimental setup was taken from \cite{SKAnoise} while we compute the baseline density as a function of visibility $n(\mathbf{u})$ by considering a Gaussian distribution of detectors with $\sigma_r = 700$m. 
We cut off the parallel modes at $j_{\rm max} = 20$ and assume a bandwidth of 5 Mhz, with 21 redshift bins ($z_{\rm min} = 6$, $z_{\rm max} = 12.4$). From Fig.~\ref{fig:SN_CMB_21} it can be seen that one would need $N^{\omega \omega(\zsource}_{\ell} \leq 10^{-14}$ in order to obtain a 3$\sigma$ rejection of the null cross. Putting aside practical issues, we estimate that a post-SKA type experiment with a baseline of 100 km (instead of $\sim 1$ km) and  the same antenna filling factor (i.e. $A_{\rm eff} = 100 A_{\rm eff}^{\rm SKA}$) would achieve such a reconstruction.  We would like to stress that the signal peaks towards higher $\zsource$, which  motivates a more detailed analysis of the signal in the redshift ranges  $15 \leq \zsource \leq 30$ ($z_{\rm max}^{\rm SKA} \sim 27$). Furthermore, lensing reconstruction relies sensitively on the smallest modes; optimal antenna distributions \cite{Koopmans2015} can be considered to measure exactly those modes. We will leave this for future work. 
\vspace{0.7em}

\noindent
{\bf Post-Reionization ($0.5 \leq \zsource \leq 2.5$):}
At the lowest redshifts we consider a CHIME-like experiment \cite{CHIME}. We assume 4 cylinders of 20 by 100 m with a total of 2048 detectors. We compute the visibility function $n(\mathbf{u})$ as in Ref.~\cite{2015ApJ...798...40X}. We assume an 80 MHz bandwidth, resulting in $\sim 5$ redshift bins in the range $0.7\leq \zsource \leq 2. 5$  We estimate the 21-cm signal as in Ref.~\cite{0004-637X-781-2-57}. Similar to lensing reconstruction using galaxies \cite{2010PhRvD..82b3522D, 2014PhRvD..90d3527C}, it would be very impractical to  detect the cross-correlation in this redshift window. The reason is that based on Fig.~\ref{fig:SN_CMB_21} even in the most optimistic case, such a probe barely reaches the 3$\sigma$ threshold for null rejection. On the other hand, it is easier to observe small scale modes  at low $\zsource$. Further investigation is needed to address whether a survey aimed at slightly higher redshifts, but with a much larger cylinder, could potentially push this towards the null rejection threshold.

\section{Conclusion and Discussion} \label{sec:conclusion}

We considered the cross-correlation between CMB $B$-modes and curl-lensing modes for sources at various redshifts. Since both are sourced by primordial gravitational waves, evidence of non-zero cross-correlation would be a strong indication that observed $B$-modes are due to primordial gravitational waves rather than astrophysical foregrounds or systematics.

We showed that, for $\ell > 2$, the cross-correlation has larger amplitude for sources at redshifts $\zsource \gtrsim 3$. This suggests that cosmic surveys aimed at mapping out higher redshifts, such as future 21-cm experiments, are much more sensitive to this cross-correlation than low redshift probes. Although the CMB provides the highest redshift screen available, the low noise required for lensing reconstruction in order to detect the cross-correlation with CMB data alone seems to be out of reach for the foreseeable future. The three-dimensional information available in 21-cm experiments allows for lower noise lensing reconstruction by combining many independent redshift slices.

We showed that in principle a post-SKA experiment aimed to detect the HI signal at redshifts $15\leq z \leq 30$ is potentially as capable of constraining this cross-correlation as much more futuristic surveys of the cosmic dark ages. 

On a side note, another motivation to investigate this cross-correlation is that it could confuse a measurement of higher order correlation functions, searching for primordial tensor non-Gaussianity \cite{Meerburg:2016ecv}. 

We are currently entering the precision CMB polarization era, with the goal of detecting the tiny signatures left by primordial gravitational waves. Before we can confidently conclude that an observation of $B$-mode polarization on large angular scales was indeed sourced by primordial gravitational waves, we need to rule out the possibility of contamination by foregrounds or systematics.  In this paper we showed that, although futuristic, the cross-correlation between CMB $B$-modes and curl lensing provides a useful test to establish the primordial nature of observed CMB $B$ modes.

\vspace{0.7em}

\noindent
{\bf Acknowledgments}
We would like to thank  Cora Dvorkin, Gil Holder,  Ue-Li Pen, and Harrison Winch for helpful discussions, and  Scott Dodelson, Marc Kamionkowski, and David Spergel for  comments on an early draft of this paper.  J.M. was supported by the Vincent and Beatrice Tremaine Fellowship.
\bibliography{BCross}

\end{document}